\documentstyle{article}
\newcommand{\Be}{\begin{equation}}
\newcommand{\Ee}{\end{equation}}
\newcommand{\Bea}{\begin{eqnarray}}
\newcommand{\Eea}{\end{eqnarray}}
\begin{document}
\begin{titlepage}

\title{A  See-Saw Mechanism with light sterile neutrinos}

\author{B.\ H.\ J.\ McKellar\thanks{e-mail b.mckellar@physics.unimelb.edu.au} 
University of Melbourne \\
Parkville, Victoria 3052, Australia\\
\\
G.\ J.\ Stephenson Jr.\ \thanks{e-mail: gjs@baryon.phys.unm.edu}\\
University of New Mexico \\
Albuquerque, New Mexico 87131 \\
\\
T.\ Goldman\thanks{e-mail goldman@t5.lanl.gov} \\
Los Alamos National Laboratory \\
Los Alamos, New Mexico 87545 \\
\\
 M.\ Garbutt\thanks{e-mail mag@physics.unimelb.edu.au} \\
University of Melbourne \\
Parkville, Victoria 3052, Australia}

\maketitle
\vspace*{-5.1in}
\flushright{LA-UR-01-3233}
\vspace{-.1in}
\flushright{UM-P-2001/018}
\vspace{-.4in}
\vspace*{4.9in}
\begin{abstract}

The usual see-saw mechanism for the generation of light neutrino masses
is based on the assumption that all of the flavours of right-handed
(more properly, sterile) neutrinos are heavy.
If the sterile Majorana mass matrix is singular, one or more of the 
sterile neutrinos will have zero mass before mixing with the active
(left-handed) neutrinos and be light after that mixing is introduced.
In particular, a rank 1 sterile mass matrix leads naturally to two
pseudo-Dirac pairs, one very light active Majorana neutrino and one
heavy sterile Majorana neutrino.  For any pattern of Dirac masses,
there exists a region of parameter space in which the two pseudo-Dirac
pairs are nearly degenerate in mass.  This, in turn, leads to large
amplitude mixing of active states as well as mixing into sterile
states.
\end{abstract}
\end{titlepage}

\section{Introduction}

Conventional wisdom holds that neutrinos ought to be Majorana particles
with very small masses, due to the action of a ``see-saw''
mechanism~\cite{see-saw}.  On the other hand, there have been recent
theoretical suggestions~\cite{GSMcK,KC} that neutrinos may well be
Majorana particles occuring in nearly degenerate pairs, the so-called
pseudo-Dirac neutrinos.  The recent studies at Super
Kamiokande~\cite{superK} of atmospheric neutrinos, which appear to
require oscillations between nearly maximally mixed mass eigenstates,
appeared to lend credence to this suggestion, although the present
analyses show that this mixing cannot be entirely to sterile
states~\cite{atmos}.

There is no overriding principle to specify the structure of the mass
matrix assumed for the sterile sector in flavor space.  Some early
discussions~\cite{wolfm2} implicitly assume that a mass term in the
sterile sector should be proportional to the unit matrix.  This has the
pleasant prospect, in terms of the initial argument for the see-saw,
that all neutrino flavors have small masses on the scale of other
fermions.  However, since there is no obvious requirement that Dirac
masses in the neutral lepton sector are the same as Dirac masses in any
other fermionic sector, this result is not compelling.  Another
possibility, which we shall discuss here, is that, in flavor space, the
rank of the mass matrix for the sterile sector is less than the number
of flavors.

In this paper, we shall concentrate on the case of a rank $1$ sterile
matrix, relegating the rank $2$ case to some remarks at the end.
Before including the effects of the sterile mass, we assume three
non-degenerate Dirac neutrinos, (although this is not essential,) which
are each constructed from one Weyl spinor which is active under the
$SU(2)_W$ of the Standard Model (SM) and one Weyl spinor which is
sterile under that interaction.  Being neutrinos, both Weyl fields have
no interactions under the $SU(3)_C$ or the $U(1)$ of the SM.  There is
then an MNS~\cite{MNS} matrix which relates these Dirac mass
eigenstates to the flavor eigenstates in the usual manner.  Note,
however, that these matrix elements are not those extracted directly
from experiment, as the mass matrix in the sterile sector will induce
additional mixing.

We next use the Dirac mass eigenstates to define bases in both the
active $3$-dimensional flavor space and the $3$-dimensional sterile
space\footnote{In much of the literature, the sterile space is
referred to as "Right-handed" (or just R) and the active space as
"Left-handed" (or just L), which follows from the behavior of the
components of a Dirac neutrino where the neutrino is defined as that
neutral lepton emitted along with a positively charged lepton.  Since
we are dealing with mass matrices, which necessarily all couple
left-handed representations to right-handed representations, we choose
to refer to these as active and sterile.}.  Following the spirit of the
original see-saw, we allow for no Majorana mass term in the active
space.

A rank $1$ sterile mass matrix may be represented as a vector of length
$M$ oriented in some direction in the $3$-dimensional sterile space.
If that vector lies along one of the axes, then that Dirac neutrino
will partake of the usual see-saw structure (one nearly sterile
Majorana neutrino with mass approximately $M$ and one nearly purely
active neutrino with mass approximately $m_D^2/M$) and the other two
mass eigenstates will remain Dirac neutrinos.  If that vector lies in a
plane perpendicular to one axis, the eigenstate associated with that
axis will remain a pure Dirac neutrino, and the other two will form one
pseudo-Dirac pair and a pair displaying the usual see-saw structure.
Both of these pairs will be mixtures of the $4$ Weyl fields associated
with the two mixing Dirac neutrinos.  In general, the structure will be
$2$ pseudo-Dirac pairs and one see-saw pair, all mixed.

As we remarked above, the very large mixing required by the atmospheric
neutrino measurements was initially taken to be evidence for a scheme
involving pseudo-Dirac neutrinos. (This, afterall, follows Pontecorvo's
initial suggestion~\cite{BP}.)  However, pure mixing into the sterile
sector is now strongly disfavored~\cite{atmos}.  It is evident from the
discusion above that there is a region of parameter space (directions
of the vector) in which the two pseudo-Dirac pairs are very nearly
degenerate, giving rise to the possibility of strong mixing in the
active sector coupled with strong mixing into the sterile sector.  We
shall explore this point in some depth.

The organization of the remainder of the paper is as follows.  In the
next section we present the mass matrix, discuss the parameterization
of the sterile mass matrix and various limiting cases.  We show the
spectrum for a general case.  In the section following that, we
specialize to the case where the vector representing the sterile mass
entry lies in a plane perpendicular to one of the axes.  In this case
we can carry out an analytical expansion in the small parameter
$<m_D>/M$.  In the fourth section we apply those results to the case
where the plane in question is perpendicular to the axis for the middle
Dirac mass, raising the possibility of near degeneracy between
pseudo-Dirac pairs.  Moving away from that plane produces large mixing
amongst the members of those pseudo-Dirac pairs.  Finally, we remark on
the structures expected for a rank $2$ sterile matrix and then
reiterate our conclusions.

\section{General mass matrix}

The flavor basis for the active neutrinos and the pairing to sterile
components defined by the (generally not diagonal) Dirac mass matrix
could be used to specify the basis for the sterile neutrino mass matrix
$M_S$. Instead we take the basis in the sterile subspace to allow the 
following convention. This implies a corresponding transformation of 
the Dirac mass matrix, which is irrelevant at present since the entries 
in that matrix are totally unkown. 

Choose a mass eigenvalue of $M$ and an eigenvector in the third
direction.  Then rotatate to Dirac mass eigenstates, first by an angle
of $\theta$ in the $1-3$ plane and then by $\phi$ in the $1-2$ plane.
This gives a $3 \times 3$ mass matrix in the sterile sector denoted by
\Be
M_S = M \left[ 
\begin{array}{ccc}
\cos^2 \phi \sin^2 \theta  &  \cos \phi \sin \phi \sin^2 \theta  &
\cos \phi \sin \theta \cos \theta  \\  \cos \phi \sin \phi \sin^2 \theta & 
\sin^2 \phi \sin^2 \theta  &  \sin \phi \sin \theta \cos \theta  \\ 
\cos \phi \sin \theta \cos \theta & \sin \phi \sin \theta \cos \theta & 
\cos^2 \theta   \end{array}  \right]. 
\Ee

In this representation, the Dirac mass matrix is 
\Be
M_D = \left[ \begin{array}{ccc}
m_1   &  0  &  0  \\  0  &  m_2  &  0  \\  0  &  0  &  m_3
\end{array} \right].  
\Ee

Note that there are special cases.  For $\theta = 0$ and any value for
$\phi$,
\Be
M_S = \left[ \begin{array}{ccc}
0 & 0 & 0 \\ 0 & 0 & 0 \\ 0 & 0 & M \end{array} \right]  .
\Ee
For $\theta = \pi / 2$ and $\phi = 0$, 
\Be
M_S = \left[ \begin{array}{ccc} 
M & 0 & 0 \\ 0 & 0 & 0 \\ 0 & 0 & 0 \end{array} \right], 
\Ee
and, for $\theta = \pi / 2$ and $\phi = \pi / 2$, 
\Be
M_S = \left[ \begin{array}{ccc} 0 & 0 & 0 \\ 0 & M & 0 \\ 0 & 0 & 0
\end{array} \right]. 
\Ee

The $6 \times 6$ submatrix\footnote{We use the states rather than the
field operators to define the mass matrix; see Ref.(\cite{SPQR}).}  of
the full $
12 \times 12$ is, in block form, 
\Be
{\cal M} = \left[ \begin{array}{cc} 0 & M_D \\ M_D & 
M_S \end{array} \right].   
\Ee

Note that, in the chiral representation, the full $12 \times 12$ matrix is
\Be
\left[ \begin{array}{cc} 0 & {\cal M} \\ 
{\cal M} & 0 \end{array} \right]. \nonumber
\Ee
Thus the full set of eigenvalues will be $\pm$ the eigenvalues of
${\cal M}$.  Where it matters for some analysis we keep track of the signs
of the eigenvalues; however for most results we present positive mass
eigenvalues.

After some algebra, we obtain the secular equation 
\Bea
0 & = & \lambda^6 -M \lambda^5 -(m_1^2 + m_2^2 + m_3^2) \lambda^4 \nonumber \\
& & + M [m_3^2 \sin^2 \theta + m_2^2 (\sin^2 \theta \cos^2 \phi + \cos^2 \theta)] 
\lambda^3 \nonumber \\ & & + (m_1^2 m_2^2 + m_2^2 m_3^2 + m_3^2 m_1^2) 
\lambda^2 \\
& & - M (m_1^2 m_2^2 \cos^2 \theta + m_2^2 m_3^2 \cos^2 \phi \sin^2 \theta 
\nonumber \\
  &  & + m_3^2 m_1^2 \sin^2 \phi \sin^2 \theta) \lambda \nonumber \\
  &  & - m_1^2 m_2^2 m_3^2. \nonumber
\Eea  

This may be rewritten as 
\Bea
 0 & = & (\lambda^2 - m_1^2) (\lambda^2 - m_2^2) 
(\lambda^2 - m_3^2) \nonumber \\ &   &    
- \lambda M \left( \lambda^4 -\left[ m_3^2 \sin^2 \theta 
+ m_2^2 (\sin^2 \theta \cos^2 \phi + \cos^2 \theta ) \right. \right. \nonumber \\
&  & \left. \left. + m_1^2 ( \sin^2 \theta \sin^2 \phi + \cos^2 \theta) 
\right] \lambda^2 \right. \\
&  & \left. +m_1^2 m_2^2 \cos^2 \theta + m_2^2 m_3^2 \sin^2 \theta 
cos^2 \phi  \right. \nonumber \\
&  & \left. + m_3^2 m_1^2 \sin^2 \theta \sin^2 \phi \right). \nonumber
\Eea

The special cases in which the sterile mass-vector is aligned along 
one of the Dirac mass matrix axes follow directly.  For $\theta = 0$, we find
\Be
(\lambda^2 - m_1^2) (\lambda^2 - m_2^2)
(\lambda^2 - M \lambda - m_3^2) = 0, 
\Ee
for $\theta = \pi / 2$ and $ \phi = 0$
\Be
(\lambda^2 - m_2^2) (\lambda^2 - m_3^2)
(\lambda^2 - M \lambda - m_1^2) = 0, 
\Ee
and for $\theta = \pi / 2$ and $\phi = \pi / 2$
\Be
(\lambda^2 - m_3^2) (\lambda^2 - m_1^2)
(\lambda^2 - M \lambda - m_2^2) = 0. 
\Ee

If $m_1^2 = m_2^2 = m_3^2 = m^2$, then we find
\Be
(\lambda^2 - m^2)^2 (\lambda^2 - M \lambda - m^2) = 0. 
\Ee

Note that in each of these cases the sterile neutrino and the original
Dirac neutrino it is aligned with become a traditional see-saw pair.

To study the system in general, we need to pick some numerical
examples.  For the current exercise, we have picked the following
parameters.

\Bea
M & = & 1000 \nonumber \\ m_1 & = & 1 \nonumber \\ 
m_2 & = & 2 \nonumber \\ m_3 & = & 3. \nonumber
\Eea

For this choice, the eigenvalues have a definite pattern for all values
of $\theta$ and $\phi$.  There are two very close pairs, with values
between $1$ and $3$.  There is one very small eigenvalue, of order
$10^{-3}$ reflecting the ratio of $m$ to $M$, and one of order
$10^{3}$ (i.e., of order $M$).  Treating the see-saw couple as a pair allows
us to present three tables of such pairs for sets of angles $\theta ,
\phi = \pi / 8, \pi / 4, 3 \pi / 8 $.

First, for the smaller close pair
\Be
\begin{array}{lccc}
\theta \backslash \phi & \pi /8  &  \pi /4  & 3 \pi / 8  \\
             &         &          &            \\
\pi / 8      & 1.398125 & 1.230175 & 1.068477   \\
             & 1.394934 & 1.228025 & 1.067688    \\
             &          &          &            \\
\pi / 4      & 1.809478 & 1.478863 & 1.151936   \\
             & 1.808183 & 1.477134 & 1.150941   \\
             &          &          &             \\
3 \pi / 8    & 1.877166 & 1.562977 & 1.18999    \\
             & 1.876742 & 1.561911 & 1.189146  \end{array}
\Ee

Then, for the next closest pair
\Be
\begin{array}{lccc}
\theta \backslash \phi & \pi / 8  &  \pi / 4 & 3 \pi / 8   \\
             &          &          &             \\
\pi / 8      & 2.038992 & 2.107688 & 2.158044    \\
             & 2.038729 & 2.107156 & 2.157407    \\
             &          &          &              \\
\pi / 4      & 2.347974 & 2.46348  & 2.529128    \\
             & 2.346047 & 2.462176 & 2.52809     \\
             &          &          &              \\
3 \pi / 8    & 2.816525 & 2.847539 & 2.868607    \\
             & 2.815691 & 2.846972 & 2.868186   
\end{array} 
\Ee

Finally, even though this pair does not directly impact the argument,
to present a complete set, we display the remaining pair
\Be
\begin{array}{lccc}
\theta \backslash \phi & \pi / 8  &  \pi / 4 & 3 \pi / 8   \\
             &          &          &             \\
\pi / 8      & 1000.008 & 1000.008 & 1000.008    \\
             & 0.00444  & 0.005366 & 0.006778    \\
             &          &          &              \\
\pi / 4      & 1000.005 & 1000.006 & 1000.006    \\
             & 0.001997 & 0.002717 & 0.004248    \\
             &          &          &              \\
3 \pi / 8    & 1000.003 & 1000.003 & 1000.004    \\
             & 0.001289 & 0.001819 & 0.003092   
\end{array} 
\Ee

\section{Two flavor subspace}

In the next section we shall discuss the case where two of the 
pseudo-Dirac pairs are nearly degenerate and follow the mixing
patterns as we move away from that region of parameter space.  To
facilitate that discussion, and to explore a system where analytic
approximations are available, it is useful to examine the limit where
one Dirac mass eigenstate remains uncoupled from the other two.  
Anticipating the following section, we decouple $m_2$.  This is 
equivalent to examining a two flavor system in which the Dirac mass
eigenvalues are $m_1$ and $m_3$ and the vector describing the sterile
mass is described by $\phi = 0$.

It is useful to define some new symbols:

\Bea
m_0^2  & = & m_1^2 \sin^2 \theta + m_3^2 \cos^2 \theta \\
a & = &\frac{ \left(m_1^2 - m_3^2\right)\sin\theta \cos
\theta}{m_0\sqrt{2}} \\
b & = & \frac{m_1m_3}{m_0}
\Eea
and $c = \cos\theta$, $s = \sin\theta$.
Note the additional $1/\sqrt{2}$ factor in $a$.

It helps to transform the mass matrix
\Be
{\cal{M}}_{1} =
\left(\begin{array}{cccc}
0 & 0 & m_{1} & 0 \\
0 & 0 & 0 & m_{3} \\
m_{1} & 0 & Ms^{2} & Mcs \\
0 & m_{3} & Mcs & Mc^{2}
\end{array}
\right)
\Ee
into the form
\Be
\cal{M} =
\left(\begin{array}{cccc}
m_0 & 0 & 0 & a \\
0 & -m_0 & 0 & -a \\
0 & 0 & 0 & b \\
a & -a & b & M
\end{array}
\right)
\Ee
so one can see that to lowest order the three small eigenvalues are
$\pm m_0, 0$.  Note the minus sign on the $a$ in the (2,4) and
(4,2) positions.  
%Omission of this sign produces the wrong type of
% shift.

The matrix effecting the transformation $ {\cal{M}} = \Omega^{\dag}
{\cal{M}}_{1} \Omega$ is
\Be
\Omega =
m_{0}^{-1}\left(\begin{array}{cccc}
m_{1}s/\sqrt{2} & -m_{1}s/\sqrt{2} & m_{3}c & 0 \\
-m_{3}c/\sqrt{2} & m_{3}c/\sqrt{2} & m_{1}s & 0 \\
m_{0}s/\sqrt{2} & m_{0}s/\sqrt{2} & 0 & m_{0}c \\
-m_{0}c/\sqrt{2} & - m_{0}c/\sqrt{2} & 0 & m_{0}c
\end{array} \right)
\Ee

This suggests writing the characteristic equation as:
\Be
\mu \left(m_{0}^{2} - \mu^{2}\right)\mu(M - \mu) =
  2 \mu^{2} a^{2} - \left(m_{0}^{2} - \mu^{2}\right) b^{2}
\Ee
which is convenient for iterative solution in a series in $M^{-1}$.
The usual equation obtained directly from $\left| \cal{M}_{1} - \mu
\right| = 0$,
\Be
\mu^{4} - \mu^{3}M - \mu^{2}\left(m_{1}^{2} + m_{3}^{2}\right) + \mu
m_{0}^{2} M + m_{1}^{2} m_{3}^{2} = 0,
\Ee
is just the same equation.

The solution to order $M^{-2}$ is now 
%(although the $M^{-2}$ terms are not yet triple checked).

\Bea
\mu_1 & = & m_0 - \frac{a^{2}}{M} -
\frac{a^{2}}{m_{0}M^{2}}\left(m_{0}^{2} - \frac{a^{2}}{2} -
b^{2}\right)\\
\mu_2 & = & -m_0 - \frac{a^{2}}{M} +
\frac{a^{2}}{m_{0}M^{2}}\left(m_{0}^{2} - \frac{a^{2}}{2} -
b^{2}\right) \\
\mu_3 & = & - \frac{b^2}{M} \\
\mu_4 & = & M + \frac{b^2}{M} + 2 \frac{a^2}{M}
\Eea

Notice that the eigenvalues sum to $M$ as they must, but the $\pm
m_{0}$ eigenvalues are shifted in the same direction at $O(M^{-1})$ and
in opposite directions at $O(M^{-2})$.  Note that $\mu_{3}$ and
$\mu_{4}$, do not pick up $O(M^{-2})$ corrections; their next
correction is at the next order.

%The solutions near $\pm m_{0}$ now specialise to the simpler case.

\section{Nearly degenerate psudo-Dirac pairs}

Applying the techniques of the last section, we find the angle
$\theta$ such that $m_0~=~m_2$.  We then move $\phi$ away from $0$ and
display the eigenfunctions by giving the amplitudes in the original
basis.  To illustrate the general nature of the result, we have
changed the Dirac masses from the even spacing used above.

In the following table, the Dirac masses are taken to be
$m_1 =  1$, $m_2 =  1.1$, and  $m_3 =  3$.  This effectively means that
$m_1$ is taken to set the scale.  To display the structure of the 
spectrum, rather than to seek a realistic example, we have
chosen $M_S = 1000$.  The angles are given in degrees.

\begin{verbatim}
__________________________________________________________________________
\end{verbatim}
$\theta =  9.324078$,   $\phi =  0$

\begin{verbatim}
      mass    1active  2active   3active   1sterile  2sterile  3sterile

   1.099328   0.635032   0.000000 -0.310533  0.698108  0.000000 -0.113793
   1.100680  -0.633620   0.000000  0.314383  0.697413  0.000000 -0.115345
   1.100000   0.000000   0.707107  0.000000  0.000000  0.707107  0.000000
   1.100000   0.000000  -0.707107  0.000000  0.000000  0.707107  0.000000
   0.007438   0.441883   0.000000  0.897064 -0.003287  0.000000 -0.002224
1000.008789   0.000162   0.000000  0.002960  0.162017  0.000000  0.986784
\end{verbatim}

\pagebreak
\noindent $\theta =  9.324078$,   $\phi =  2.25$

\begin{verbatim}
    mass      1active   2active   3active  1sterile  2sterile  3sterile

   1.095953   0.479130   0.468214  -0.225940  0.525106 -0.466489 -0.082539
   1.096608   0.437964  -0.514829  -0.208027 -0.480274  0.513243  0.076041
   1.103359   0.416946   0.529767  -0.212981  0.460041  0.531383 -0.078333
   1.104056  -0.458049  -0.484588   0.235669  0.505710  0.486376 -0.086730
   0.007438   0.441553   0.015769   0.897088 -0.003285 -0.000109 -0.002224
1000.008789   0.000162   0.000007   0.002960  0.161892  0.006361  0.986784
\end{verbatim}
\noindent $\theta =  9.324078$,   $\phi =  4.5$

\begin{verbatim}
    mass      1active   2active   3active  1sterile  2sterile  3sterile

   1.092254  0.479875 -0.471453 -0.217491  0.524155 -0.468127 -0.079183
   1.092888 -0.458763  0.495815  0.209390  0.501371 -0.492614 -0.076279
   1.107010  0.416602  0.526536 -0.221472  0.461189  0.529886 -0.081725
   1.107726  0.437718  0.503654 -0.234323 -0.484866 -0.507196  0.086521
   0.007439  0.440571  0.031517  0.897156 -0.003273 -0.000217 -0.002226
1000.008789  0.000162  0.000014  0.002960  0.161517  0.012712  0.986784
\end{verbatim}
\noindent $\theta =  9.324078$   $\phi =  22.5$

\begin{verbatim}
    mass      1active   2active   3active  1sterile  2sterile  3sterile

    1.062925  0.550356 -0.405921 -0.179528  0.584987  -0.392239 -0.063608
    1.063381  0.546548 -0.411257 -0.179609 -0.581185   0.397574  0.063663
    1.134871  0.337840  0.568726 -0.249457  0.383405   0.586755 -0.094367
    1.135731  0.341702  0.564710 -0.254038 -0.388074  -0.583058  0.096172
    0.007475  0.409265  0.154109  0.899298 -0.003058  -0.001048 -0.002241
 1000.008789  0.000150  0.000068  0.002960  0.149684   0.062001  0.986784
\end{verbatim}
\noindent $\theta =  9.324078$   $\phi =  45$

\begin{verbatim}
     mass    1active   2active   3active   1sterile  2sterile  3sterile

   1.030458  0.632073 -0.290233 -0.127244  0.651329 -0.271878 -0.043708
   1.030692 -0.630801  0.292859  0.127989  0.650162 -0.274406 -0.043972
   1.163620  0.226485  0.612428 -0.270973  0.263544  0.647849 -0.105102
   1.164612  0.227955  0.610618 -0.274587 -0.265472 -0.646488  0.106595
   0.007566  0.315141  0.286490  0.904762 -0.002384 -0.001969 -0.002282
1000.008789  0.000115  0.000126  0.002960  0.114563  0.114563  0.986784
________________________________________________________________________
\end{verbatim}

This table represents a small part of the available parameter space and
was chosen to display some possible features.  First, $\theta$ was 
chosen so that, at $\phi = 0$, the Dirac pair at $m_2$ was bracketed
by the pseudo-Dirac pair.  Such a $\theta$ will exist for any pattern 
of the Dirac masses.  Then, for small values of $\phi$, there will be
two nearly degenerate pseudo-Dirac pairs.

Note that, for $\phi = 0$, there is no mixing between the fields labelled
by $2$ and the remaining fields, while for the next entry at $\phi = 2.25$
degrees there is considerable mixing.  That mixing continues as $\phi$
increases while the eigenvalues move apart.  The pattern described by the centroids of the pseudo-Dirac pairs is fixed by the angles $\theta$ and
$\phi$.  If $M_S$ is increased, that pattern is hardly changed.  The 
primary effect of increasing $M_S$, as would be expected from the analysis
in the previous section, is to decrease the separation of the two members
of each pseudo-Dirac pair while producing the usual see-saw behavior for
the remaining pair.

The implication for oscillation phenomena is clear.  A given weak
interaction produces an active flavor eigenstate which is some linear combination of the three active components listed in the table.  That then
translates into a linear combination of the six mass eigenstates.  From the
table it is clear that the involvement of the heavy Majorana see-saw state
is minimal, so the expansion really consists of the light Majorana see-saw
state and the four Majorana states arising from the two pseudo-Dirac pairs.

Since these five mass eigenstates have both active and sterile components,
the subsequent time evolution will involve both flavor change and
oscillation into the sterile sector.

Finally, inspection of the column labelled 1active for $\phi = 2.25$ or
$\phi = 4.5$, for example, shows that the presence of a rank $1$ sterile
mass matrix can seriously change any mixing pattern of the MNS 
type~\cite{MNS}, from that which would have obtained with purely Dirac
neutrinos.

\section{Rank $2$}

We have not discussed the case of rank $2$ matrices explicitly,
although the pattern is obvious.  In such a case, there would be two
see-saw pairs and one pseudo-Dirac pair, leading to three active and
one sterile light neutrino.  While this pattern is currently being
analyzed in the literature, we do not see any compelling pattern for
it in the sterile sector.  Furthermore, the current interpretation of
the atmospheric neutrino data can be accomodated much more easily in
the rank $1$ case discussed in this paper.  Therefore we leave the
discussion of rank $2$ to a longer paper.

\section{Conclusions}

Our analysis shows that, starting with a rank 1 mass matrix in the 
sterile sector, it is possible to generate in a simple way, scenarios in 
which one of the active neutrinos mixes in a near maximal way with both 
active and sterile neutrinos.  The super-Kamiokande data allow such 
mixing.

We are encouraged by this to pursue the concepts presented here in 
more detail in a later paper. 

\section{Acknowledgments}
This research is supported in part by the Department of Energy under
contract W-7405-ENG-36, in part by the Australian Research Council and in
part by the National Science Foundation.


\begin{thebibliography}{99}
\bibitem{see-saw} M.\ Gell-Mann, P.\ Ramond and R.\ Slansky, in {\it
Supergravity}, Proceedings of the Workshop, Stony Brook, New York,
1979, edited by P.\ van Nieuwenhuizen and D.\ Freedman (North-Holland,
Amsterdam 1979, p. 315; T.\ Yanagida, in {\it Proceedings of the
Workshop on the Unified Theories  and Baryon Number in the Universe},
Tsukuba, Japan, 1979, edited by O.\ Sawada and A.\ Sugamoto (KEK Report
No. 79-18, Tsukuba, 1979), p.95.
\bibitem{GSMcK} T.\ Goldman, G.\ J.\ Stephenson Jr.\ and
B.\ H.\ J.\ McKellar, Mod. Phys. Lett. {\bf A15}, (2000) 439.
\bibitem{KC} Kevin Cahill, hep-ph/9912416 (1999); hep-ph/9912508 (1999).
\bibitem{superK} Y.\ Fukuda {\it et al.} {\it Phys. Rev. Lett.} {\bf
81} (1998) 1562; {\bf 82} (1999) 2644; Phys. Lett. B {\bf 476} (1999)
185. 
\bibitem{atmos} S.\ Fukuda {\it et al.} Phys. Rev. Lett. {\bf 85}, (2000) 
3999; G.\ L.\ Folgi, E.\ Lisi and A.\ Marrone, Phys. Rev. {\bf D63},
(2001) 053008; A.\ De Rujula, M.\ B.\ Gavela and P.\ Hernandez, Phys. Rev.
{\bf D63}, (2001) 033001.
\bibitem{wolfm2} L.\ Wolfenstein, Phys. Lett. {\bf B107}, (1981) 77.
\bibitem{MNS} Z.\ Maki, M.\ Nakagawa and S.\ Sakata, Prog. Theor. Phys.
{\bf 28}, (1962) 870.
\bibitem{BP} B.\ Pontecorvo, Zh. Eskp. Teor. Fiz. {\bf 33}. (1957) 549
[Sov. Phys. JETP {\bf 6}, (1958) 479]; Zh. Eksp. Teor. Fiz. {\bf 34},
(1958) 247.
\bibitem{SPQR} see, {\it e.g.} W.\ C.\ Haxton and G.\ J.\ Stephenson, Jr.,
Progress in Particle and Nuclear Physics (Sir Denys Wilkinson, editor)
Vol. {\bf 12}, Permagon Press, New York, (1984) 409; S.\ M.\ Bilenky and 
S.\ T.\ Petcov, Rev. Mod. Phys. {\bf 59}, (1987) 671.

\end{thebibliography}
\end{document}